\documentclass[universe,conferenceproceedings,submit,pdftex,moreauthors]{Definitions/mdpi}

\usepackage{amsthm}
\newtheorem{thm}{Theorem}
\newtheorem*{cor}{Corollary}
\usepackage{wasysym}
\nolinenumbers 

\firstpage{1} 
\pubvolume{1}
\issuenum{1}
\articlenumber{0}
\pubyear{2022}
\copyrightyear{2022}

\Title{The Scale Invariant Vacuum Paradigm: \\ main results and current progress}

\Author{Vesselin G. Gueorguiev $^{1,2,\dagger,}$*\orcidV{} and Andre Maeder $^{3}$\orcidA{} }

\AuthorNames{Vesselin G. Gueorguiev and Andre Maeder}

\address{%
$^{1}$ \quad Institute for Advanced Physical Studies, Sofia 1784, Bulgaria;\\
$^{2}$ \quad Ronin Institute for Independent Scholarship, 127 Haddon Pl., Montclair, NJ 07043, USA\\
$^{3}$ \quad Geneva Observatory, University of Geneva, Chemin des Maillettes 51, CH-1290 Sauverny, Switzerland; 
Andre.Maeder at unige.ch 
}

\corres{Correspondence: Vesselin at MailAPS dot org}

\firstnote{Presenter at the AlteCosmoFun'21 on Sep. 9th, 2021} 
\secondnote{This paper is a summary of the talk presented at the conference: 
Alternative Gravities and Fundamental Cosmology,
Organized by the University of Szczecin, Poland, 6 -10 September 2021.}

\nolinenumbers 

\abstract{
We present a summary of the main results within the Scale Invariant Vacuum (SIV) paradigm 
as related to the Weyl Integrable Geometry. After a brief review of the mathematical framework,
we will highlight the main results related to inflation within the SIV \citep{SIV-Inflation'21}, 
the growth of the density fluctuations \citep{MaedGueor19}, 
and the application of the SIV to scale-invariant dynamics of Galaxies, MOND, 
Dark Matter, and the Dwarf Spheroidals \citep{MaedGueor20b}. 
The connection of the weak-field SIV results to the un-proper time parametrization within 
the reparametrization paradigm is also discussed \citep{2021Symm...13..379G}.
}


\begin{document}
\nolinenumbers 

\section{Motivation}
\subsection{Scale Invariance and Physical Reality}

The presence of a scale is related to the existence of physical connection and causality.
The corresponding relationships are formulated as physical laws dressed in mathematical expressions.
The laws of physics (formulae) change upon change of scale, as a result, 
using consistent units is paramount in physics and leads to powerful dimensional estimates
on the order of magnitude of physical quantities. 
The underlined scale is closely related to the presence of material content.

However, in the absence of matter a scale is not easy to define. 
Therefore, an empty Universe would be expected to be scale invariant!
Absence of scale is confirmed by the scale invariance of the 
Maxwell equations in vacuum (no charges and no currents - the sources of the electromagnetic fields).
The field equations of General Relativity are scale invariant for empty space with zero cosmological constant.
What amount of matter is sufficient to kill scale invariance is still an open question.
Such question is particularly relevant to Cosmology and the evolution of the Universe.

\subsection{Einstein GR (EGR) and Weyl Integrable Geometry (WIG). }

Einstein's General Relativity (EGR) is based on the premise of a torsion-free covariant connection 
that is metric-compatible and guarantees preservation of the length of vectors along geodesics
$(\delta\left\Vert \overrightarrow{v}\right\Vert =0)$.
The theory has been successfully tested at various scales, starting from local Earth laboratories,
the Solar system, on galactic scales via light-bending effects, 
and even on an extragalactic level via the observation of gravitational waves.
The EGR is also the foundation for modern Cosmology and Astrophysics. 
However, at galactic and cosmic scales, some new and mysterious 
phenomena have appeared. The explanations for these phenomena are often 
attributed to unknown matter particles or fields that are yet to be detected 
in our laboratories - Dark Matter and Dark Energy. 
Since these new particles and/or fields have evaded any laboratory detection 
for more than twenty years then it seems plausible to turn 
to the other alternative - a modification of EGR.

In the light of the above discussion one may naturally ask could 
the mysterious phenomena be artifacts of non-zero
$\delta\left\Vert \overrightarrow{v}\right\Vert $, but often negligible 
and with almost zero value $(\delta\left\Vert \overrightarrow{v}\right\Vert \approx0)$,
which could accumulate over cosmic distances and fool us that the 
observed phenomena may be due to Dark Matter and/or Dark Energy?
An idea of extension of EGR was proposed by Weyl as soon as GR was proposed by Einstein. 
Weyl proposed an extension to GR by adding local gauge (scale) invariance 
that does have the consequence that lengths may not be preserved upon parallel transport.
However, it was quickly argued that such model will result in path dependent phenomenon 
and thus contradicting observations. A cure was later found to this objection 
by introducing the Weyl Integrable Geometry (WIG) where the lengths of vectors  
are conserved only along closed paths ($\varoint\delta\left\Vert \overrightarrow{v}\right\Vert =0$).
Such formulation of the Weyl's original idea defeats the Einstein objection!
Even more, given that all we observe about the distant Universe are waves
that get to us then the condition for Weyl Integrable Geometry is basically saying 
that the information that arrives to us via different paths is interfering constructively 
to build a consistent picture of the source object. 

One way to build a WIG model is to consider conformal transformation 
of the metric field $g'_{\mu\nu}=\lambda^{2}g_{\mu\nu}$ and to apply it
to various observational phenomena. As we will see in the discussion below
the demand for homogenous and isotropic space restricts the field $\lambda$ 
to depend only on the cosmic time and not on the space coordinates.
The weak field limit of such WIG model results in an extra acceleration in the
equation of motion that is proportional to the velocity of the particle.
Even more, the Scale Invariant Vacuum (SIV) idea provides a way of finding out the
specific functional form of $\lambda(t)$ as applicable to LFRW cosmology
and its WIG extension. 

We also find it important to point out that extra acceleration in the
equations of motion, which is proportional to the velocity of a particle, 
could also be justified by requiring re-parametrization symmetry.
Not implementing re-parametrization invariance in a model could lead
to un-proper time parametrization \cite{2021Symm...13..379G} 
that seems to induce ``fictitious forces'' in the equations of motion 
similar to the forces derived in the weak field SIV regime. It is a 
puzzling observation that may help us understand nature better.

\section{Mathematical Framework}

\noindent
\subsection{Weyl Integrable Geometry and Dirac co-calculus}

The  framework for the Scale Invariant Vacuum paradigm 
is based on the Weyl Integrable Geometry and Dirac co-calculus as mathematical tools for 
description of nature \cite{Weyl23,Dirac73}. 
The original Weyl Geometry uses a metric tensor field $g_{\mu\nu}$, 
along with a ``connexion'' vector field $\kappa_{\mu}$, and a scalar field $\lambda$.
In the Weyl Integrable Geometry the ``connexion'' vector field $\kappa_{\mu}$ is not an independent 
field but it is derivable from the scalar field $\lambda$.

\begin{equation}
\kappa_{\mu}=-\partial_{\mu}\ln(\lambda)
\label{connexion}
\end{equation}
This form of the ``connexion'' vector field $\kappa_{\mu}$ guarantees its irrelevance, 
in the covariant derivatives, upon integration over closed paths.
That is, $\varoint \kappa_{\mu}dx^{\mu} =0$. In other words, $\kappa_{\mu}dx^{\mu}$ represents a closed 1-form, 
and even more, it is an exact form since \eqref{connexion} implies $\kappa_{\mu}dx^{\mu}=d\lambda$.
Thus, the scalar function $\lambda$ plays a key role in the Weyl Integrable Geometry.
Its physical meaning is related to the freedom of a local scale gauge that provides 
a description upon change in scale via local re-scaling $l'\rightarrow\lambda(x)l$.

\subsection{Gauge change \& derivatives. EGR and WIG frames.}
The covariant derivatives use the rules of the Dirac co-calculus  \cite{Dirac73} 
where tensors also have co-tensor powers based on the way they transform upon change of scale.
For the metric tensor $g_{\mu\nu}$ this power is $n=2$. 
This follows from  the way the length of a line segment $ds$ 
with coordinates $dx^{\mu}$ is defined via the usual expression $ds^2=g_{\mu\nu}dx^{\mu}dx^{\nu}$. 

$$l'\rightarrow\lambda(x)l\Leftrightarrow ds'=\lambda ds \Rightarrow g'_{\mu\nu}=\lambda^{2}g_{\mu\nu}$$
This leads to $g^{\mu\nu}$ having the co-tensor power of $n=-2$ in order to have the 
Kronecker $\delta$ as scale invariant object ($g_{\mu\nu}g^{\nu\rho}=\delta_{\mu}^{\rho}$).
That is, a co-tensor is of power $n$ when upon local scale change it satisfies:

\begin{equation}
l'\rightarrow\lambda(x)l :\;  Y'_{\mu\nu}\rightarrow\lambda^{n}Y_{\mu\nu}
\label{co-tensor}
\end{equation}
In the Dirac co-calculus this results in the appearance of the ``connexion'' vector field $\kappa_{\mu}$ 
in the  covariant derivatives of scalars, vectors, and tensors:

\begin{table}[h]
\caption{Derivatives for co-tensors of power n.}
\begin{center}
\begin{tabular}{|rl|}
\hline
co-tensor type & mathematical expression \\
\hline
co-scalar & $S_{*\mu}=\partial_{\mu}S-n\kappa_{\mu}S$,\\
co-vector & $A_{\nu*\mu}=\partial_{\mu}A_{\nu}-\;^{*}\Gamma_{\nu\mu}^{\alpha}A_{\alpha}-n\kappa_{\nu}A_{\mu}$,\\
co-covector & $A_{*\mu}^{\nu}=\partial_{\mu}A^{\nu}+\;^{*}\Gamma_{\mu\alpha}^{\nu}A^{\alpha}-nk^{\nu}A_{\mu}$.\\
\hline
\end{tabular}
\end{center}
\label{Table1}
\end{table}%

\noindent 
where the usual Christoffel symbol $\Gamma_{\mu\alpha}^{\nu}$ is replaced by 
$^{*}\Gamma_{\mu\alpha}^{\nu}=\Gamma_{\mu\alpha}^{\nu}+g_{\mu\alpha}k^{\nu}-g_{\mu}^{\nu}\kappa_{\alpha}-g_{\alpha}^{\nu}\kappa_{\mu}$.

The corresponding equation of the geodesics within the WIG was first introduced by Dirac in 1973 \cite{Dirac73} ($u^{\mu}=dx^{\mu}/{ds}$):
\begin{equation}
u_{*\nu}^{\mu}=0\Rightarrow\frac{du^{\mu}}{ds}
+^{*}\varGamma_{\nu\rho}^{\mu}u^{\nu}u^{\rho}
+\kappa_{\nu}u^{\nu}u^{\mu}=0\,.\label{eq:geodesics}
\end{equation}

\noindent 
This geodesic equation has also been derived from reparametrisation-invariant action by Bouvier \& Maeder in 1978:
\begin{equation*}
\delta\mathcal{A}=\intop_{P_{0}}^{P_{1}}\delta\left(d\widetilde{s}\right)
=\int\delta\left(\beta ds\right)=\int\delta\left(\beta\frac{ds}{d\tau}\right)d\tau=0.\label{eq:action}
\end{equation*}

\subsection{Scale Invariant Cosmology}

The scale invariant cosmology equations were first introduced in \citeyear{Dirac73} by \citet{Dirac73},
and then re-derived in \citeyear{Canuto77} by \citet{Canuto77}. The equations are based on the
corresponding expressions of the Ricci tensor and the relevant extension of the Einstein equations.

\subsubsection{The Einstein equation for Weyl's geometry}
The conformal transformation ($g'_{\mu\nu}=\lambda^{2}g_{\mu\nu}$)  
of the metric tensor $g_{\mu\nu}$ in the more general Weyl's frame into
Einstein frame, where the metric tensor is $g'_{\mu\nu}$, induces a simple relation between 
the Ricci tensor and scalar in Weyl's Integrable Geometry and the Einstein GR frame 
(using prime to denotes Einstein GR frame objects):
\begin{eqnarray*}
R_{\mu\nu}=R'_{\mu\nu}-\kappa_{\mu;\nu}-\kappa_{\nu;\mu}-2\kappa_{\mu}\kappa_{\nu}
+2g_{\mu\nu}\kappa^{\alpha}\kappa_{\alpha}-g_{\mu\nu}\kappa_{;\alpha}^{\alpha}\,,\\
R=R'+6\kappa^{\alpha}\kappa_{\alpha}-6\kappa_{;\alpha}^{\alpha}\,.
\end{eqnarray*}

\noindent
When considering the Einstein equation along with the above expressions, one obtains:
\begin{eqnarray}
R_{\mu\nu}-\frac{1}{2}\ g_{\mu\nu}R=-8\pi GT_{\mu\nu}-\Lambda\,g_{\mu\nu}\,,\\
R'_{\mu\nu}-\frac{1}{2}\ g_{\mu\nu}R'-\kappa_{\mu;\nu}-\kappa_{\nu;\mu}
-2\kappa_{\mu}\kappa_{\nu}+2g_{\mu\nu}\kappa_{;\alpha}^{\alpha}-g_{\mu\nu}\kappa^{\alpha}\kappa_{\alpha}=\nonumber \\
-8\pi GT_{\mu\nu}-\Lambda\,g_{\mu\nu}\,.
\label{field}
\end{eqnarray}

The relationship $\Lambda=\lambda^{2}\Lambda_{\mathrm{E}}$ of $\Lambda$ in WIG
to the Einstein cosmological constant $\Lambda_{\mathrm{E}}$ in the EGR
was present in the original form of the equations to provide explicit scale invariance.
This relationship makes explicit the appearance of $\Lambda_{\mathrm{E}}$ as invariant scalar (in-scalar) 
since then one has $\Lambda\,g_{\mu\nu}=\lambda^{2}\Lambda_{\mathrm{E}}\,g_{\mu\nu}=\Lambda_{\mathrm{E}}\,g'_{\mu\nu}$.

The above equations are a generalization of the original Einstein GR equation.
Thus, they have even a larger class of local gauge symmetries that need to be fixed by a gauge choice.
In Dirac's work the gauge choice was based on the large numbers hypothesis.
Here we discuss a different gauge choice.

The corresponding scale-invariant FLRW based cosmology equations 
within the WIG frame were first introduced in \citeyear{Canuto77} by \citet{Canuto77}:
\begin{eqnarray}
\frac{8\,\pi G\varrho}{3}=\frac{k}{a^{2}}+\frac{\dot{a}^{2}}{a^{2}}+2\,\frac{\dot{\lambda}\,\dot{a}}{\lambda\,a}
+\frac{\dot{\lambda}^{2}}{\lambda^{2}}-\frac{\Lambda_{\mathrm{E}}\lambda^{2}}{3}\,,\label{E1p}\\
-8\,\pi Gp=\frac{k}{a^{2}}+2\frac{\ddot{a}}{a}+2\frac{\ddot{\lambda}}{\lambda}+\frac{\dot{a}^{2}}{a^{2}}
+4\frac{\dot{a}\,\dot{\lambda}}{a\,\lambda}-\frac{\dot{\lambda^{2}}}{\lambda^{2}}-\Lambda_{\mathrm{E}}\,\lambda^{2}\,.\label{E2p}
\end{eqnarray}
These equations clearly reproduce the standard FLRW equations in the limit $\lambda=const=1$.
The scaling of $\Lambda$ was recently used to revisit the 
Cosmological Constant Problem within quantum cosmology \cite{GueorM20}.
The conclusion of \cite{GueorM20} is that our Universe is unusually large 
given that the expected mean size of all Universes, 
where Einstein GR holds, is expected to be of a Plank scale.
In the study, $\lambda=const$ was a key assumption since the 
Universes were expected to obey the Einstein GR equations.
It is an open question what would be the expected mean size of all Universes
if the condition  $\lambda=const$ is relaxed as for a WIG-Universes ensemble.

\subsubsection{The Scale Invariant Vacuum gauge ($T=0$ and $R'=0$)} 

The idea of the Scale Invariant Vacuum was introduced first in \citeyear{Maeder17a} by \citet{Maeder17a}.
It is based on the fact that for Ricci flat ($R'_{\mu\nu}=0$) Einstein GR vacuum ($T_{\mu\nu}=0$)
one obtains form \eqref{field} the following equation for the vacuum:
\begin{equation}
\kappa_{\mu;\nu}+\kappa_{\nu;\mu}+2\kappa_{\mu}\kappa_{\nu}
-2g_{\mu\nu}\kappa_{;\alpha}^{\alpha}+g_{\mu\nu}\kappa^{\alpha}\kappa_{\alpha}=\Lambda\,g_{\mu\nu}\label{SIV}
\end{equation}

For homogeneous and isotropic WIG-space $\partial_{i}\lambda=0$;
therefore, only $\kappa_{0}=-\dot{\lambda}/\lambda$ and its time
derivative $\dot{\kappa}_{0}=-\kappa_{0}^{2}$ can be non-zero.
As a corollary of \eqref{SIV} one can derive the following set of equations \cite{Maeder17a}:
\begin{eqnarray}
\ 3\,\frac{\dot{\lambda}^{2}}{\lambda^{2}}\,=\Lambda\,,\quad\mathrm{and}\quad2\frac{\ddot{\lambda}}{\lambda}
-\frac{\dot{\lambda}^{2}}{\lambda^{2}}\,=\Lambda\,,\label{SIV1}\\
\mathrm{or}\quad\frac{\ddot{\lambda}}{\lambda}\,=\,2\,\frac{\dot{\lambda}^{2}}{\lambda^{2}}\,,
\quad\mathrm{and}\quad\frac{\ddot{\lambda}}{\lambda}-\frac{\dot{\lambda}^{2}}{\lambda^{2}}\,=\frac{\Lambda}{3}\,.\label{SIV2}
\end{eqnarray}
One could derive these equations by using the time and space components of the equations or 
by looking at the relevant trace invariant along with the relationship $\dot{\kappa}_{0}=-\kappa_{0}^{2}$.
Any pair of these equations is  sufficient to prove the other pair of equations. 

\begin{thm}
Using the SIV equations \eqref{SIV1} or \eqref{SIV2} with $\Lambda=\lambda^{2}\Lambda_{E}$ one has: 
\begin{equation}
\Lambda_{E}=3\frac{\dot{\lambda^{2}}}{\lambda^{4}},\quad\mathrm{with}\quad\frac{d\Lambda_{E}}{dt}=0.
\label{SIV-gauge}
\end{equation}
\end{thm}
 
\begin{cor} 
The solution of the SIV equations is: $\lambda=t_{0}/t$, 
and $t_0=\sqrt{3/\Lambda_{E}}$ with $c=1$.
\end{cor}

Upon the use of the SIV gauge, first done in  \citeyear{Maeder17a} by \citet{Maeder17a}, one observes that 
{\bf the cosmological constant disappears} from the equations \eqref{E1p} and \eqref{E2p}:
\begin{eqnarray}
\frac{8\,\pi G\varrho}{3}=\frac{k}{a^{2}}+\frac{\dot{a}^{2}}{a^{2}}+2\,\frac{\dot{a}\dot{\lambda}}{a\lambda}\,,\label{E1}\\
-8\,\pi Gp=\frac{k}{a^{2}}+2\frac{\ddot{a}}{a}+\frac{\dot{a^{2}}}{a^{2}}+4\frac{\dot{a}\dot{\lambda}}{a\lambda}\,.\label{E2}
\end{eqnarray}

\section{Comparisons and Applications}

The predictions and outcomes of the SIV paradigm were confronted 
with observations in series of papers of the current authors.
Highlighting the main results and outcomes is the subject of current section.

\subsection{Comparing the scale factor $a(t)$ within  $\Lambda$CDM and SIV.}

\begin{figure}[h]
\centering 
\includegraphics[width=0.8\textwidth]{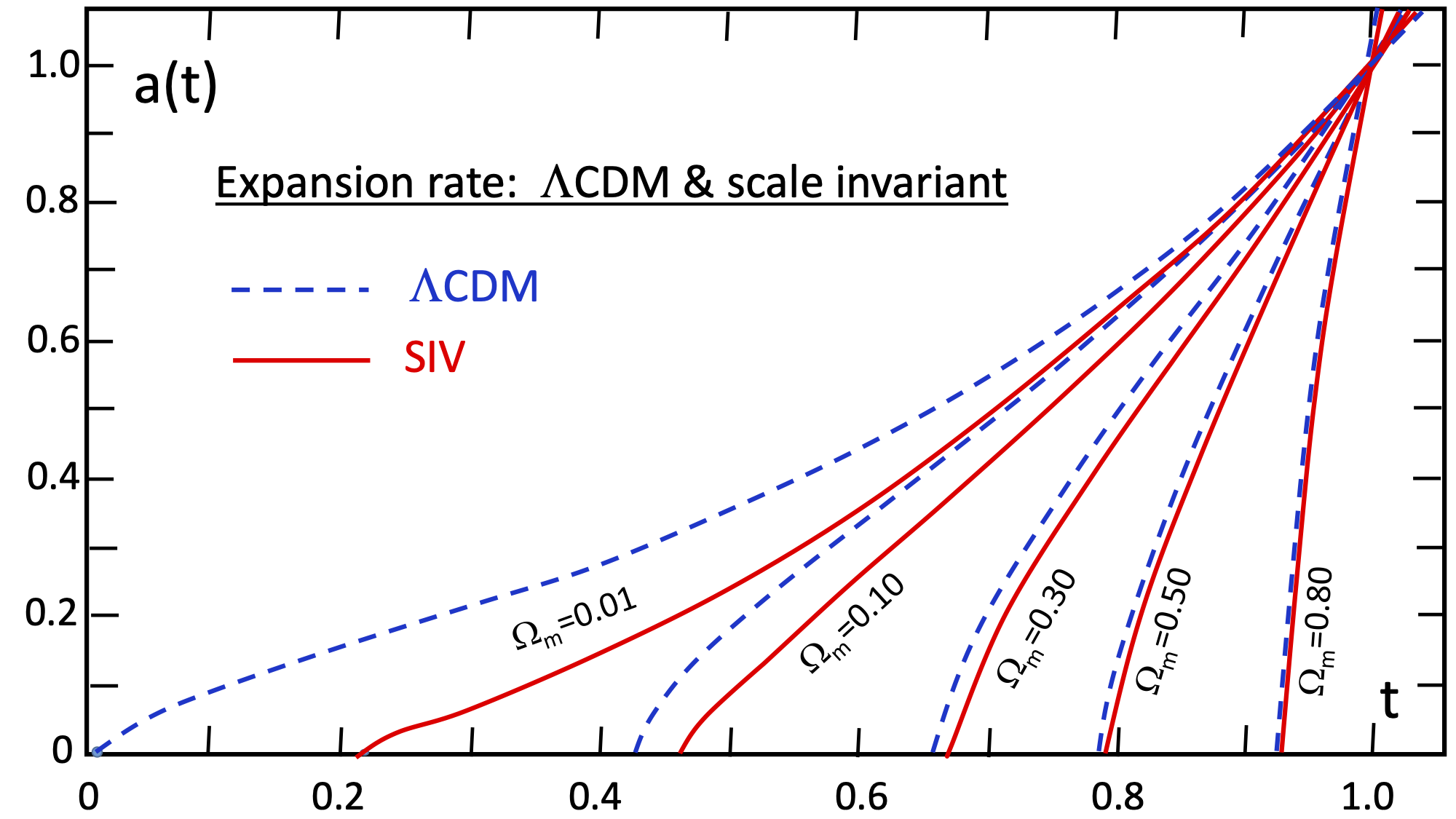} 
\caption{
Expansion rates $a(t)$ as a function of time $t$ in the flat
($k=0$) $\Lambda$CDM and SIV models in the matter dominated era.
The curves are labeled by the values of $\Omega_{\mathrm{m}}$.}
\label{rates} 
\end{figure}

Upon arriving at the SIV cosmology equations \eqref{E1} and \eqref{E2}, 
along with the gauge fixing \eqref{SIV-gauge},  which implies $\lambda=t_0/t$ 
with $t_0$ - the current age of the Universe since the Big-Bang ($a=0$ and $t=0$), 
the implications for cosmology were first discussed by \citet{Maeder17a} and 
later reviewed by \citet{MaedGueor20a}. The most important point in 
comparing $\Lambda$CDM and SIV cosmology models is the existence of 
SIV cosmology with slightly different parameters but almost the same 
curve for the standard scale parameter $a(t)$ 
when the time scale is set so that  $t_0=1$ now 
\cite{Maeder17a,MaedGueor20a}. 
As seen in Fig. \ref{rates} the differences between the 
$\Lambda$CDM and SIV models 
declines for increasing matter densities.

\subsection{Application to Scale-Invariant Dynamics of Galaxies}

The next important application of the scale-invariance at cosmic scales is
the derivation of a universal expression for the
Radial Acceleration Relation (RAR) of $g_{\mathrm{obs}}$  and $g_{\mathrm{bar}}$.
That is, the relation between the observed gravitational acceleration $g_{\mathrm{obs}}=v^2/r$
and the acceleration from the baryonic matter due to the 
standard Newtonian gravity $g_\mathrm{N}$ by \cite{MaedGueor20b}:

\begin{equation}
g\,=\,g_{\mathrm{N}}+\frac{k^{2}}{2}+\frac{1}{2}\sqrt{4g_{\mathrm{N}}k^{2}+k^{4}}\,,
\label{sol}
\end{equation}
where $g=g_{\mathrm{obs}}$, $g_N=g_{\mathrm{bar}}$.
For $g_{\mathrm{N}} \gg k^{2} : g \rightarrow g_{\mathrm{N}}$
but for $g_{\mathrm{N}}\rightarrow 0 \Rightarrow g \rightarrow k^{2}$ is a constant.

\begin{figure}[h]
\centering 
\includegraphics[width=0.75\textwidth]{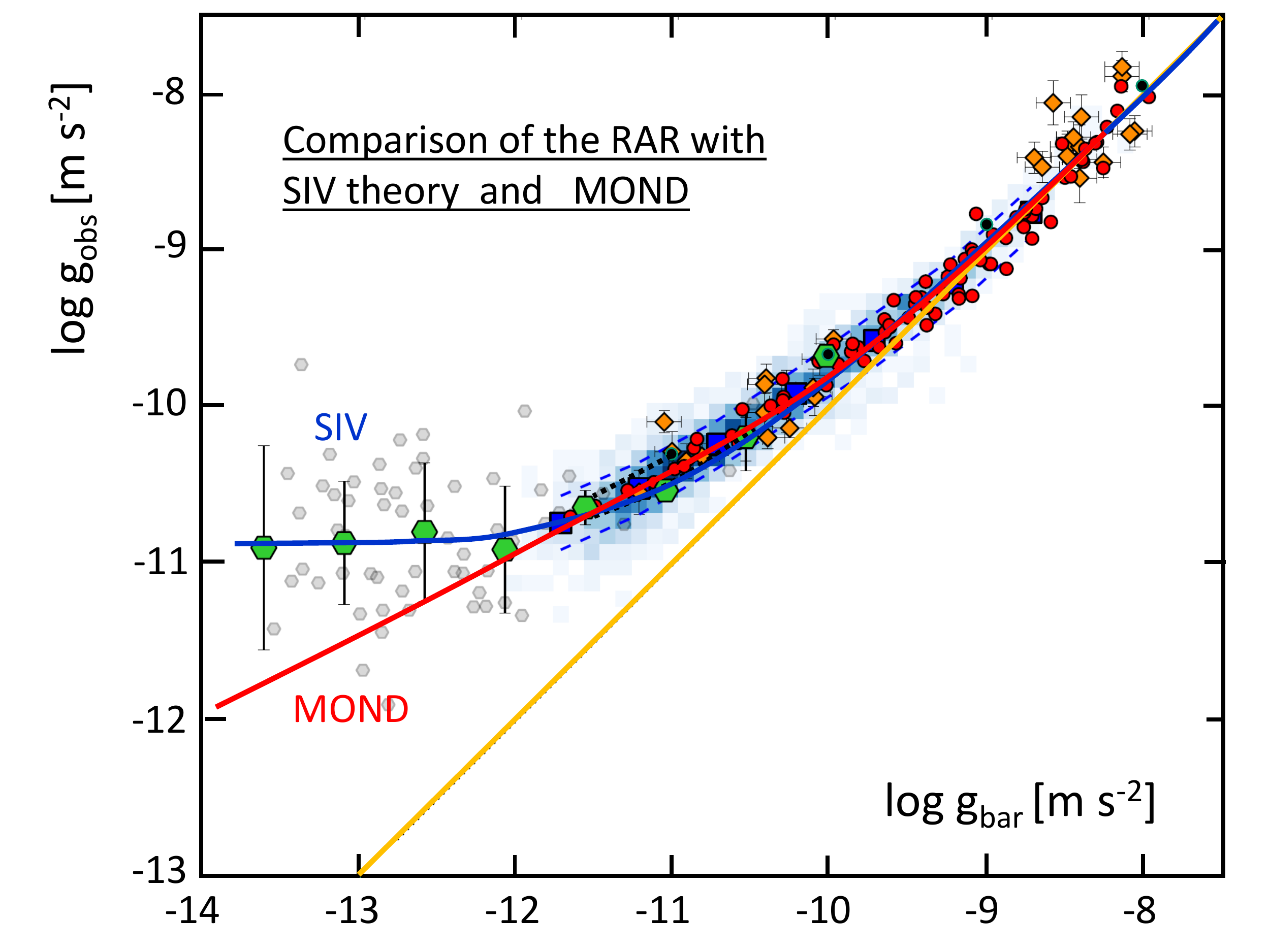} 
\caption{
Radial Acceleration Relation (RAR) for the galaxies studied by Lelli et al. (2017).
Dwarf Spheroidals  as binned data (big green hexagons), 
along with MOND (red curve) and SIV (blue curve) model predictions.
The orange curve shows the 1:1-line for $g_{\mathrm{obs}}$  and $g_{\mathrm{bar}}$.}
\label{gobs} 
\end{figure}

As seen in Fig. \ref{gobs} MOND deviates significantly for the data on the Dwarf Spheroidals.
This is well-known problem in MOND due to the need of two different interpolating functions,
one in galaxies and one at cosmic scales. The SIV universal expression \eqref{sol} resolves 
this issue naturally with one universal parameter $k^2$ related to the gravity at large distances 
\cite{MaedGueor20b}. 

The expression \eqref{sol} follows from the Weak Field Approximation (WFA) of the SIV upon
utilization of the Dirac co-calculus in the derivation of the geodesic equation within the 
relevant WIG \eqref{eq:geodesics}, see \cite{MaedGueor20b} for more details:  
\begin{eqnarray}
g_{ii}=-1,\;g_{00}&=&1+2\Phi/c^{2}\Rightarrow\varGamma_{00}^{i}
=\frac{1}{2}\frac{\partial g_{00}}{\partial x^{i}}=\frac{1}{c^{2}}\frac{\partial\Phi}{\partial x^{i}},\nonumber\\
\frac{d^{2}\overrightarrow{r}}{dt^{2}}&=&-\frac{GM}{r^{2}}\frac{\overrightarrow{r}}{r}
+\kappa_{0}(t)\frac{d\overrightarrow{r}}{dt}.\label{Nvec}
\end{eqnarray}
where $i\in{1,2,3}$, while the potential $\Phi=GM/r$ is scale invariant.

By considering the scale-invariant ratio of the correction term $\kappa_0(t) \, \upsilon\, $ 
to the usual Newtonian term in \eqref{Nvec}, one has:
\begin{equation}
x=\frac{\kappa_0 \upsilon r^{2}}{GM}=\frac{H_{0}}{\xi}\frac{\upsilon \,r^{2}}{GM}
= \, \frac{H_0}{\xi}  \frac{(r \, g_{\mathrm{obs}})^{1/2}}{g_{\mathrm{bar}}}
\sim\frac{g_{\mathrm{obs}}-g_{\mathrm{bar}}}{g_{\mathrm{bar}}}\,,
\end{equation}

Then by utilizing an explicit scale invariance for cancelling the proportionality factor:
\begin{equation}
{\left(\frac{g_{\mathrm{obs}}-g_{\mathrm{bar}}}{g_{\mathrm{bar}}}\right)_{2}}\div
{\left(\frac{g_{\mathrm{obs}}-g_{\mathrm{bar}}}{g_{\mathrm{bar}}}\right)_{1}}\,=
\,\left(\frac{g_{\mathrm{obs, 2}}}{ g_{\mathrm{obs, 1}}}\right)^{1/2}\,
\left(\frac{g_{\mathrm{bar,1}}}{g_{\mathrm{bar,2}}}\right)\,,
\label{corrg}
\end{equation}
by setting $g=g_{\mathrm{obs, 2}}$, $g_N=g_{\mathrm{bar,2}}$, 
and with $k=k_{(1)}$ all the system-1 terms, one has:
\begin{equation*}
\frac{g}{g_{\mathrm{N}}}-1=k_{(1)}\frac{g^{1/2}}{g_{\mathrm{N}}}
\Rightarrow
g\,=\,g_{\mathrm{N}}+\frac{k^{2}}{2}\pm\frac{1}{2}\sqrt{4g_{\mathrm{N}}k^{2}+k^{4}}.
\end{equation*}

\subsection{Growth of the Density Fluctuations within the SIV}

Another interesting result was the possibility of a fast growth of the density fluctuations 
within the SIV \citep{MaedGueor19}. This study modifies accordingly 
the relevant equations like continuity equation, Poisson equation, and Euler equation 
within the SIV framework. Here we outline the main equations and the relevant results.

By using the highlighted notation ${\color{blue}\kappa}=\kappa_0=-\dot{\lambda}/\lambda=1/t$, 
the corresponding Continuity, Poisson, and Euler equations are:
\begin{eqnarray*}
\frac{\partial \rho}{\partial t}+ \vec{\nabla}\cdot (\rho \vec{v}) 
= \color{blue}\kappa\color{black} \left [\rho+ \vec{r} \cdot \vec{\nabla} \rho \right]\,,\; 
\vec{\nabla}^{2}\Phi=\triangle\Phi=4\pi G \varrho \label{eq:Continuity+Poisson},\\
\frac{d\vec{v}}{dt}=\frac{\partial\vec{v}}{\partial t}+\left(\vec{v}\cdot\vec{\nabla}\right)\vec{v}=
-\vec{\nabla}\Phi-\frac{1}{\rho}\vec{\nabla}p+\color{blue}\kappa\color{black}\vec{v}\label{Euler} \, .
\end{eqnarray*}

For a density perturbation $\varrho(\vec{x},t)\, = \, \varrho_{b}(t)(1+\delta(\vec{x},t))$
the above equations result in:
\begin{eqnarray}
\dot{\delta} + \vec{\nabla} \cdot \dot{\vec{x}} =\kappa \vec{x} \cdot \vec{\nabla} \delta = n \kappa(t) \delta &,&
\vec{\nabla}^{2}\Psi = 4\pi Ga^{2}\varrho_{b}\delta,\label{D1}\\
\ddot{\vec{x}}+ 2 H \dot{\vec{x}}+ (\dot{\vec{x}}\cdot \vec{\nabla}) \dot{\vec{x}} 
&=& -\frac{\vec{\nabla} \Psi}{a^2} +\kappa(t) \dot{\vec{x}}.\\
\Rightarrow \ddot{\delta}+ (2H -(1+n) \color{blue}\kappa\color{black} )\dot{\delta} 
&=& 4\pi G \varrho_{b}\delta + 2 n \color{blue}\kappa\color{black} (H-\color{blue}\kappa\color{black}) \delta.\;\;
\label{D2}
\end{eqnarray}

\begin{figure}[h]
\centering 
\includegraphics[width=0.6\textwidth]{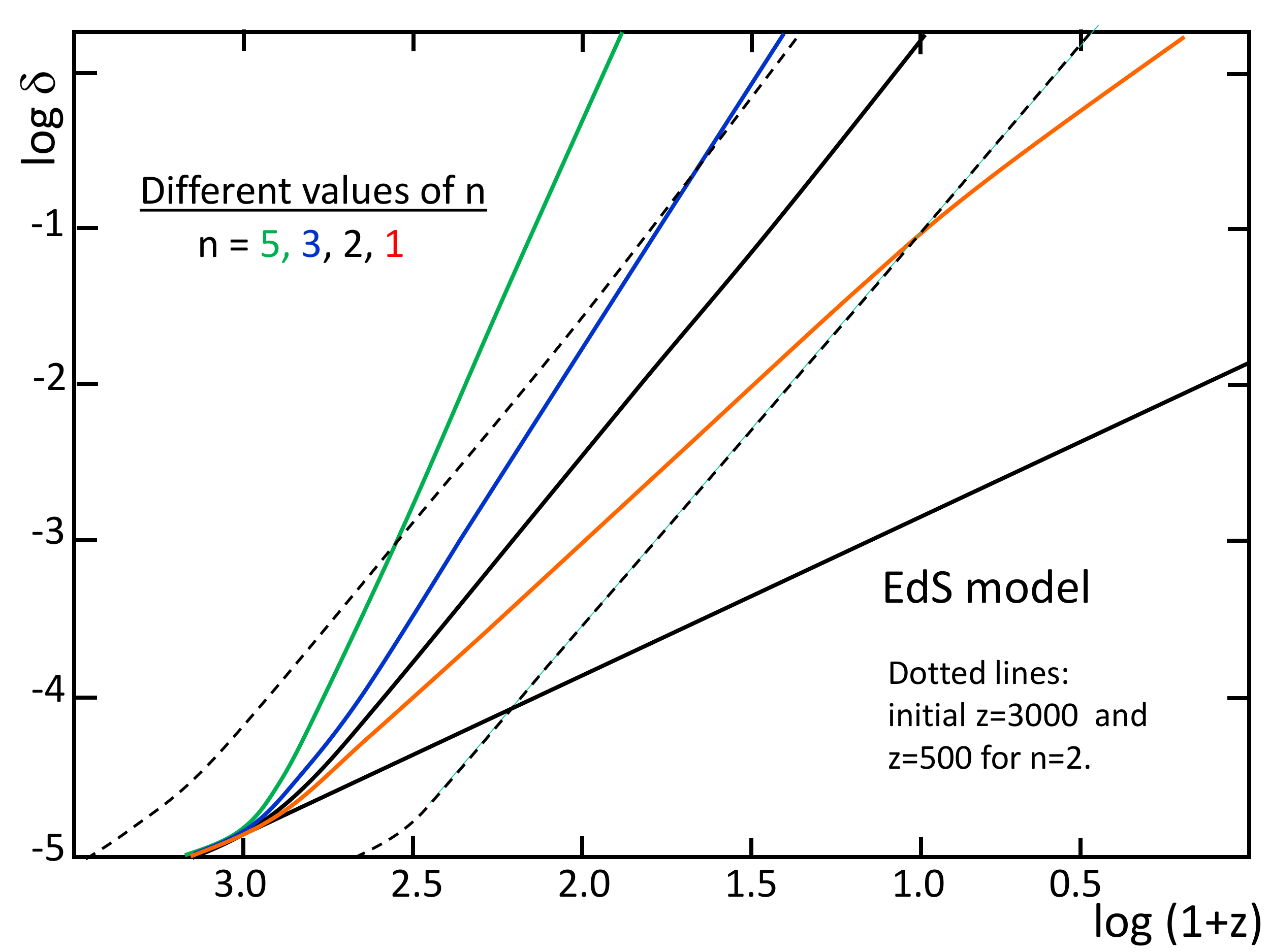} 
\caption{
The growth of density fluctuations for different values of parameter
$n$ (the gradient of the density distribution in the nascent cluster),
for an initial value $\delta=10^{-5}$ at $z=1376$ and $\Omega_{\mathrm{m}}=0.10$.
The initial slopes are those of the EdS models. The two light broken
curves show models with initial $(z+1)=3000$ and 500, with same $\Omega_{\mathrm{m}}=0.10$
and $n=2$. These dashed lines are to be compared to the black continuous
line of the $n=2$ model. All the three lines for $n=2$ are very
similar and nearly parallel.}
\label{variousn} 
\end{figure}

The final result \eqref{D2} recovers the standard equation 
in the limit of $ \color{blue}\kappa\color{black} \rightarrow 0$.
The simplifying assumption $\vec{x} \cdot \vec{\nabla} \delta(x) = n \delta(x)$ 
in \eqref{D1} introduces the parameter $n$ that measures the perturbation
type (shape). For example, a spherically symmetric perturbation would have $n=2$.
As seen in Fig. \ref{variousn} perturbations for various values of $n$ are 
resulting in faster growth of the density fluctuations within the SIV then 
the Einstein -- de Sitter model even at relatively law matter densities. 
Furthermore, the overall slope is independent of the choice of recombination epoch $z_\mathrm{rec}$.
The behavior for different $\Omega_m$  is also interesting and is shown and discussed in details
by \citet{MaedGueor19}. 

\subsection{SIV and the Inflation of the Early Universe.}

The latest result within the SIV paradigm is the presence of inflation stage at the very early Universe
$t\approx 0$ with a natural exit from inflation in a later time $t_\mathrm{exit}$ with 
value related to the parameters of the inflationary potential  \citep{SIV-Inflation'21}.
The main steps towards these results are outlined below.

If we go back to the general scale-invariant cosmology equation \eqref{E1p},
we can identify a vacuum energy density expression that relates 
the Einstein cosmological constant with the energy density
as expressed in terms of $\kappa=-\dot{\lambda}/\lambda$ by using the SIV result 
\eqref{SIV-gauge}. The corresponding vacuum energy density $\rho$, with $C=3/(4\pi G)$, is then:
\[
\rho=\frac{\Lambda}{8\pi G}=\lambda^{2}\rho'=\lambda^{2}\frac{\Lambda_{E}}{8\pi G}
=\frac{3}{8\pi G}\frac{\dot{\lambda}^{2}}{\lambda^{2}}
=\frac{C}{2}\dot{\psi}^{2}\,.
\]
This provides a natural connection to inflation within the SIV 
via $\dot{\psi}=-\dot{\lambda}/\lambda$ or $\psi\propto\ln(t)$. 
The equations for the energy-density, pressure, and Weinberg's condition for inflation 
within the standard model for inflation \citep{Guth81,Linde95,Linde05,Weinberg08} are:
\begin{eqnarray}
\left.\begin{array}{c}\rho\\ p \end{array}\right\} 
=\frac{1}{2}\dot{\varphi}^{2}\pm V(\varphi),\label{rp}\;
\mid\dot{H}_{\mathrm{infl}}\mid\,\ll H_{\mathrm{infl}}^{2}\,.
\label{cond}
\end{eqnarray}

\noindent 
If we make the identification between the standard model for inflation above with the 
fields present within the SIV (using $C=3/(4\pi G)$):
\begin{eqnarray}
\dot{\psi}=-\dot{\lambda}/\lambda, & \varphi\leftrightarrow\sqrt{C}\,\psi, \quad
V\leftrightarrow CU(\psi), & U(\psi)\,=\,g\,e^{\mu\,\psi}\,.
\label{SIV-identification}
\end{eqnarray}
Here $U(\psi)$ is the inflation potential with strength $g$ and field ``coupling'' $\mu$. 
One can evaluate the Weinberg's condition for inflation \eqref{cond} 
within the SIV framework \citep{SIV-Inflation'21}, and the result is:

\begin{equation}
\frac{\mid\dot{H}_{\mathrm{infl}}\mid}{H_{\mathrm{infl}}^{2}}\,
=\,\frac{3\,(\mu+1)}{g\,(\mu+2)}\,t^{-\mu-2}\ll1\,{\normalcolor for}\ \mu<-2,\ {\normalcolor and}\ t\ll t_{0}=1.
\label{crit}
\end{equation}

From the expression \eqref{crit} one can see that there is a graceful exit from inflation at 
a later time $t_\mathrm{exit}\approx\sqrt[n]{\frac{n\ g}{3(n+1)}}$ with $n=-\mu-2>0$,
when the Weinberg's condition for inflation \eqref{cond} is not satisfied anymore.

The derivation of the equation \eqref{crit} starts with the use of 
the scale invariant energy conservation equation within SIV \citep{Maeder17a,SIV-Inflation'21}:
\begin{equation}
\frac{d(\varrho a^{3})}{da}+3\,pa^{2}+(\varrho+3\,p)\frac{a^{3}}{\lambda}\frac{d\lambda}{da}=0\,,
\label{conserv}
\end{equation}
which has the following equivalent form:
\begin{equation}
\dot{\varrho}+3\,\frac{\dot{a}}{a}\,(\varrho+p)+\frac{\dot{\lambda}}{\lambda}\,(\varrho+3p)\,=\,0\,.
\label{conserv2}
\end{equation}

By substituting the expressions for $\rho$ and $p$ from \eqref{rp} 
along with the SIV identification \eqref{SIV-identification} 
within the SIV expression (\ref{conserv2}), one obtains modified form of the Klein-Gordon equation,
which could be non-linear when using non-linear potential $U(\psi)$ as in \eqref{SIV-identification}:
\begin{equation}
\ddot{\psi}+U'\,+3H_{\mathrm{infl}}\,\dot{\psi}-2\,(\dot{\psi}^{2}-U)\,=\,0\,.
\label{cons1}
\end{equation}

The above equation \eqref{cons1} can be used to evaluate the time derivative of the Hubble parameter.
The process is utilizing \eqref{SIV-gauge}; that is, 
$\lambda=t_{0}/t,\;\dot{\psi}=-\dot{\lambda}/\lambda=1/t\;\Rightarrow \ddot{\psi}=-\dot{\psi}^{2}$
along with $\psi=\ln(t)+const$ and $U(\psi)\,=\,g\,e^{\mu\,\psi}=g t^{\mu}$ when the normalization of the 
field $\psi$ is chosen so that $\psi(t_0)=\ln(t_0)=0$ for $t_0=1$ at the current epoch. The final result is: 
\begin{eqnarray}
H_{\mathrm{infl}}=\dot{\psi}-\frac{2\,U}{3\,\dot{\psi}}-\frac{U'}{3\,\dot{\psi}} &=&\,\frac{1}{t}-\frac{(2+\mu)\,g}{3}\,t^{\mu+1}\,,
\label{hh}\\
\dot{H}_{\mathrm{infl}}=-\dot{\psi}^{2}-\frac{2U}{3}-U'-\frac{U''}{3} &=&-\frac{1}{t^{2}}-\frac{(\mu+2)(\mu+1)\,g}{3}t^{\mu}\,.
\label{hd}
\end{eqnarray}

For $\mu<-2$ the $t^{\mu}$ terms above are dominant; 
thus, the critical ratio \eqref{cond} for the occurrence of inflation near $t\approx 0$ is then:
\begin{equation*}
\frac{\mid\dot{H}_{\mathrm{infl}}\mid}{H_{\mathrm{infl}}^{2}}\,=\,\frac{3\,(\mu+1)}{g\,(\mu+2)}\,t^{-\mu-2}\,.
\end{equation*}


\section{Conclusions and Outlook}

From the highlighted results in the previous section on various comparisons and potential applications, 
we see that the \textit{SIV cosmology is a viable alternative to $\Lambda$CDM.} 
In particular, within the SIV gauge (\ref{E1}) \textit{the cosmological constant disappears}.
There are diminishing differences in the values of the scale factor $a(t)$ 
within $\Lambda$CDM and SIV at higher densities as emphasized in the 
discussion of (Fig.~\ref{rates}) \cite{Maeder17a,MaedGueor20a}. 
Furthermore, the SIV also shows consistency for $H_0$ and the age of the Universe,  
and the m-z diagram is well satisfied - see \citet{MaedGueor20a} for details.

Even more, \textit{the SIV provides the correct RAR for dwarf spheroidals} (Fig.~\ref{gobs}) 
while MOND is failing and dark matter cannot account for the phenomenon \cite{MaedGueor20b}.  
Furthermore, it seems that  \textit{within the SIV dark matter is not needed to seed the growth of structure} 
in the Universe since there is a fast enough growth of the density fluctuations as seen in (Fig.\ref{variousn}) 
and discussed in more details by \citet{MaedGueor19}.

In our latest studies on the inflation within the SIV cosmology \cite{SIV-Inflation'21},
we have identified a connection of the scale factor $\lambda$, and its rate of change, 
with the inflation field $\psi \rightarrow \varphi\,,\; \dot{\psi}=-\dot{\lambda}/\lambda$ \eqref{SIV-identification}.
As seen from \eqref{crit} \textit{inflation of the very-very early Universe ($t\approx0$) 
is natural and SIV predicts a graceful exit from inflation}!

Some of the obvious future research directions are related to the primordial nucleosynthesis,
where preliminary results show a satisfactory comparison between SIV and observations \cite{nucleosynthesis}. 
The recent success of the R-MOND in the description of the CMB \cite{2021PhRvL.127p1302S}, 
after the initial hope and concerns \cite{Skordis2006}, is very stimulating and demands testing SIV cosmology 
against the MOND and $\Lambda$CDM successes in the description of the CMB.

Other less obvious research directions are related to exploration of SIV within the solar system due to the 
high-accuracy data available. Or exploring further and in more details the possible connection of SIV 
with the re-parametrization invariance. For example, it is already known by \citet{2021Symm...13..379G} 
that un-proper time parametrization can lead to SIV like equation of motion \eqref{eq:geodesics}.

\authorcontributions{
Conceptualization, V. Gueorguiev and A. Maeder; 
Writing – original draft, V. Gueorguiev; 
Writing – review \& editing, V. Gueorguiev and A. Maeder.}

\acknowledgments{
A.M. expresses his gratitude to his wife for her patience and support.
V.G. is extremely grateful to his wife and daughters for their understanding and family support  
during the various stages of the research presented.
This research did not receive any specific grant from funding agencies in the public, commercial, or not-for-profit sectors.
}


\reftitle{References}



\begin{thebibliography}{999}


\bibitem[Weyl(1923)]{Weyl23} {Weyl}, H. 1923, 
\textit{Raum, Zeit, Materie.}
Vorlesungen {ü}ber allgemeine Relativit{ä}tstheorie. 
Re-edited by Springer, Berlin, Heidelberg, Berlin (1993).
\href{https://doi.org/10.1007/978-3-642-78365-4}{eBook ISBN: \color{blue}978-3-642-78365-4}.

\bibitem[{Dirac(1973)}]{Dirac73} Dirac, P.~A.~M. 
{\em Proc. R. Soc. Lond. A} {\bf 1973}, {\em 333}, 403.
\href{https://www.jstor.org/stable/78370}{JStor:  \color{blue}78370}.

\bibitem[{{Canuto} {et~al.}(1977){Canuto}, {Adams}, {Hsieh}, & {Tsiang}}]{Canuto77}
{Canuto}, V., {Adams}, P.~J., {Hsieh}, S.-H., \& {Tsiang}, E.,
{\em Phys. Rev. D} {\bf 1977}, {\em 16}, 1643.
\href{https://doi.org/10.1103/PhysRevD.16.1643}{DOI:  \color{blue}10.1103/PhysRevD.16.1643}.

\bibitem[Gueorguiev  \& Maeder(2020)]{GueorM20} 
Gueorguiev, V., Maeder, A. 
{\em Universe}, {\bf 2020}, {\em 5}, 108.
\textit{Revisiting the Cosmological Constant Problem within Quantum Cosmology.}
\href{https://doi.org/10.3390/universe6080108}{ DOI:  \color{blue}10.3390/universe6080108}.


\bibitem[Maeder(2017)]{Maeder17a} Maeder, A.
{\em Astrophys. J.} {\bf 2017}, {\em 834}, 194.
\textit{An Alternative to the LambdaCDM Model: the case of scale invariance,}
\href{https://arxiv.org/abs/1701.03964}{e-Print: \color{blue}1701.03964 [astro-ph.CO].}

\bibitem[Maeder and Gueorguiev(2020)]{MaedGueor20a} 
Maeder, A.; Gueorguiev, V.G. 2020, 
{\em Universe} {\bf 2020}, {\em 6}, 46. 
\textit{The Scale-Invariant Vacuum (SIV) Theory: 
A Possible Origin of Dark Matter and Dark Energy,} 
\href{https://www.mdpi.com/2218-1997/6/3/46}{ DOI:  \color{blue}10.3390/universe6030046}.

\bibitem[Maeder and Gueorguiev(2020)]{MaedGueor20b} 
Maeder, A.; Gueorguiev, V.G. 
{\em MNRAS} {\bf 2019}, {\em 492}, 2698.
\textit{Scale-invariant dynamics of galaxies, MOND, dark matter, and the dwarf spheroidals,}
\href{https://arxiv.org/abs/2001.04978}{e-Print: \color{blue}2001.04978 [gr-qc]}

\bibitem[Maeder and Gueorguiev(2019)]{MaedGueor19} 
Maeder, A., Gueorguiev, V., G.,
{\em Phys. Dark Univ.} {\bf 2019}, {\em 25}, 100315. 
\textit{The growth of the density fluctuations in the scale-invariant vacuum theory,}
\href{https://arxiv.org/abs/1811.03495}{e-Print: \color{blue}1811.03495 [astro-ph.CO]}

\bibitem[Maeder and Gueorguiev(2021)]{SIV-Inflation'21}
Maeder, A., Gueorguiev, V.~G.,
{\em MNRAS} {\bf 2021}, {\em 504}, 4005.
\textit{Scale invariance, horizons, and inflation.}
\href{https://arxiv.org/abs/2104.09314}{e-Print: \color{blue}2104.09314 [gr-qc].}

\bibitem[Guth(1981)]{Guth81} Guth, A.
{\em Phys. Rev. D} {\bf 1981}, {\em 23}, 347.
\href{https://doi.org/10.1103/PhysRevD.23.347}{ DOI:  \color{blue}10.1103/PhysRevD.23.347}.

\bibitem[Linde(1995)]{Linde95} Linde, A. 
Lectures on Inflationary Cosmology
in Particle Physics and Cosmology - Proceedings of the Ninth Lake Louise Winter Institute, 1995;
\href{https://arxiv.org/abs/hep-th/9410082}{e-Print: \color{blue}9410082 [hep-th].}

\bibitem[Linde(2005)]{Linde05} Linde, A. , 
{\em Contemp. Concepts Phys.} {\bf 2005}, {\em 5}, pp1-362;
\textit{Particle Physics and Inflationary Universe},
\href{https://arxiv.org/abs/hep-th/0503203}{e-Print: \color{blue}0503203 [hep-th]}.

\bibitem[Weinberg(2008)]{Weinberg08} 
Weinberg, S. \textit{ Cosmology}, Oxford Univ. Press, Oxford, UK, 2008; p. 593;
\href{https://global.oup.com/academic/product/cosmology-9780198526827}{ISBN:  \color{blue}9780198526827}.

\bibitem[Gueorguiev and Maeder(2021)]{2021Symm...13..379G} 
Gueorguiev, V.~G., Maeder, A., 
{\em Symmetry} {\bf 2021}, {\em 13}, 379;
\textit{Geometric Justification of the Fundamental Interaction Fields for the Classical Long-Range Forces.} 
\href{https://arxiv.org/abs/1907.05248}{e-Print: \color{blue}1907.05248 [math-ph]}.

\bibitem[Maeder(2019)]{nucleosynthesis}
Maeder, A.
\textit{Evolution of the early Universe in the scale invariant theory.}
\href{https://arxiv.org/abs/1902.10115}{e-Print: \color{blue}1902.10115 [astro-ph.CO]}.

\bibitem[Skordis and Z{\l}o{\'s}nik(2021)]{2021PhRvL.127p1302S} 
Skordis, C., Z{\l}o{\'s}nik, T., 
{\em Phys. Rev. Lett.} {\bf 2021} {\em 127}, 1302;
New Relativistic Theory for Modified Newtonian Dynamics.
doi:10.1103/PhysRevLett.127.161302; 
\href{https://arxiv.org/abs/2007.00082}{e-Print: \color{blue}2007.00082 [astro-ph.CO]}.

\bibitem[Skordis et al.(2006)]{Skordis2006}
C. Skordis, D. F. Mota, P. G. Ferreira, and C. Boehm
{\em Phys. Rev. Lett.} {\bf 2006}, {\em 96}, 11301
Large Scale Structure in Bekenstein’s Theory of Relativistic Modified Newtonian Dynamics,
\href{https://arxiv.org/abs/astro-ph/0505519}{e-Print: \color{blue}0505519 [astro-ph]}.

\end{thebibliography}


\end{document}